\documentstyle[twoside,epsfig]{article}

\catcode`\@=11
\long\def\@makefntext#1{
\protect\noindent \hbox to 3.2pt {\hskip-.9pt  
$^{{\eightrm\@thefnmark}}$\hfil}#1\hfill}               %CAN BE USED 

\def\thefootnote{\fnsymbol{footnote}}
\def\@makefnmark{\hbox to 0pt{$^{\@thefnmark}$\hss}}    %ORIGINAL 
        
\def\ps@myheadings{\let\@mkboth\@gobbletwo
\def\@oddhead{\hbox{}
\rightmark\hfil\eightrm\thepage}   
\def\@oddfoot{}\def\@evenhead{\eightrm\thepage\hfil
\leftmark\hbox{}}\def\@evenfoot{}
\def\sectionmark##1{}\def\subsectionmark##1{}}

\oddsidemargin=\evensidemargin
\renewcommand{\thefootnote}{\fnsymbol{footnote}}

\newcounter{sectionc}\newcounter{subsectionc}\newcounter{subsubsectionc}
\renewcommand{\section}[1] {\vspace{12pt}\addtocounter{sectionc}{1} 
\setcounter{subsectionc}{0}\setcounter{subsubsectionc}{0}\noindent 
        {\tenbf\thesectionc. #1}\par\vspace{5pt}}
\renewcommand{\subsection}[1] {\vspace{12pt}\addtocounter{subsectionc}{1} 
        \setcounter{subsubsectionc}{0}\noindent 
        {\bf\thesectionc.\thesubsectionc. {\kern1pt \bfit #1}}\par\vspace{5pt}}
\renewcommand{\subsubsection}[1] {\vspace{12pt}\addtocounter{subsubsectionc}{1}
        \noindent{\tenrm\thesectionc.\thesubsectionc.\thesubsubsectionc.
        {\kern1pt \tenit #1}}\par\vspace{5pt}}
\newcommand{\nonumsection}[1] {\vspace{12pt}\noindent{\tenbf #1}
        \par\vspace{5pt}}

\newcounter{appendixc}
\newcounter{subappendixc}[appendixc]
\newcounter{subsubappendixc}[subappendixc]
\renewcommand{\thesubappendixc}{\Alph{appendixc}.\arabic{subappendixc}}
\renewcommand{\thesubsubappendixc}
        {\Alph{appendixc}.\arabic{subappendixc}.\arabic{subsubappendixc}}

\renewcommand{\appendix}[1] {\vspace{12pt}
        \refstepcounter{appendixc}
        \setcounter{figure}{0}
        \setcounter{table}{0}
        \setcounter{lemma}{0}
        \setcounter{theorem}{0}
        \setcounter{corollary}{0}
        \setcounter{definition}{0}
        \setcounter{equation}{0}
        \renewcommand{\thefigure}{\Alph{appendixc}.\arabic{figure}}
        \renewcommand{\thetable}{\Alph{appendixc}.\arabic{table}}
        \renewcommand{\theappendixc}{\Alph{appendixc}}
        \renewcommand{\thelemma}{\Alph{appendixc}.\arabic{lemma}}
        \renewcommand{\thetheorem}{\Alph{appendixc}.\arabic{theorem}}
        \renewcommand{\thedefinition}{\Alph{appendixc}.\arabic{definition}}
        \renewcommand{\thecorollary}{\Alph{appendixc}.\arabic{corollary}}
        \noindent{\tenbf Appendix \theappendixc #1}\par\vspace{5pt}}
\newcommand{\subappendix}[1] {\vspace{12pt}
        \refstepcounter{subappendixc}
        \noindent{\bf Appendix \thesubappendixc. {\kern1pt \bfit #1}}
        \par\vspace{5pt}}
\newcommand{\subsubappendix}[1] {\vspace{12pt}
        \refstepcounter{subsubappendixc}
        \noindent{\rm Appendix \thesubsubappendixc. {\kern1pt \tenit #1}}
        \par\vspace{5pt}}

\newcommand{\textlineskip}{\baselineskip=13pt}
\newcommand{\smalllineskip}{\baselineskip=10pt}

\def\eightcirc{
\begin{picture}(0,0)
\put(4.4,1.8){\circle{6.5}}
\end{picture}}
\def\eightcopyright{\eightcirc\kern2.7pt\hbox{\eightrm c}} 

\newcommand{\copyrightheading}[1]
        {\vspace*{-2.5cm}\smalllineskip{\flushleft
        {\footnotesize International Journal of Modern Physics C #1}\\
        {\footnotesize $\eightcopyright$\, World Scientific Publishing
         Company}\\
         }}

\newcommand{\publisher}[2]{{\begin{center}\footnotesize\smalllineskip 
        Received #1\\
        Revised #2
        \end{center}
        }}

\def\abstracts#1#2#3{{
        \centering{\begin{minipage}{4.5in}\footnotesize\baselineskip=10pt
        \parindent=0pt #1\par 
        \parindent=15pt #2\par
        \parindent=15pt #3
        \end{minipage}}\par}} 

\def\keywords#1{{
        \centering{\begin{minipage}{4.5in}\footnotesize\baselineskip=10pt
        {\footnotesize\it Keywords}\/: #1
        \end{minipage}}\par}}

\newcommand{\bibit}{\nineit}
\newcommand{\bibbf}{\ninebf}
\renewenvironment{thebibliography}[1]
        {\frenchspacing
         \ninerm\baselineskip=11pt
         \begin{list}{\arabic{enumi}.}
        {\usecounter{enumi}\setlength{\parsep}{0pt}     
         \setlength{\leftmargin 12.7pt}{\rightmargin 0pt} %FOR 1--9 ITEMS
         \setlength{\itemsep}{0pt} \settowidth
        {\labelwidth}{#1.}\sloppy}}{\end{list}}

\newcounter{itemlistc}
\newcounter{romanlistc}
\newcounter{alphlistc}
\newcounter{arabiclistc}

\newcommand{\fcaption}[1]{
        \refstepcounter{figure}
        \setbox\@tempboxa = \hbox{\footnotesize Fig.~\thefigure. #1}
        \ifdim \wd\@tempboxa > 5in
           {\begin{center}
        \parbox{5in}{\footnotesize\smalllineskip Fig.~\thefigure. #1}
            \end{center}}
        \else
             {\begin{center}
             {\footnotesize Fig.~\thefigure. #1}
              \end{center}}
        \fi}

\newcommand{\tcaption}[1]{
        \refstepcounter{table}
        \setbox\@tempboxa = \hbox{\footnotesize Table~\thetable. #1}
        \ifdim \wd\@tempboxa > 5in
           {\begin{center}
        \parbox{5in}{\footnotesize\smalllineskip Table~\thetable. #1}
            \end{center}}
        \else
             {\begin{center}
             {\footnotesize Table~\thetable. #1}
              \end{center}}
        \fi}

\def\@citex[#1]#2{\if@filesw\immediate\write\@auxout
        {\string\citation{#2}}\fi
\def\@citea{}\@cite{\@for\@citeb:=#2\do
        {\@citea\def\@citea{,}\@ifundefined
        {b@\@citeb}{{\bf ?}\@warning
        {Citation `\@citeb' on page \thepage \space undefined}}
        {\csname b@\@citeb\endcsname}}}{#1}}

\newif\if@cghi
\def\cite{\@cghitrue\@ifnextchar [{\@tempswatrue
        \@citex}{\@tempswafalse\@citex[]}}
\def\citelow{\@cghifalse\@ifnextchar [{\@tempswatrue
        \@citex}{\@tempswafalse\@citex[]}}
\def\@cite#1#2{{$\null^{#1}$\if@tempswa\typeout
        {IJCGA warning: optional citation argument 
        ignored: `#2'} \fi}}

\def\pmb#1{\setbox0=\hbox{#1}
        \kern-.025em\copy0\kern-\wd0
        \kern.05em\copy0\kern-\wd0
        \kern-.025em\raise.0433em\box0}

\def\fnt#1#2{\footnotetext{\kern-.3em
        {$^{\mbox{\scriptsize #1}}$}{#2}}}

\def\ps@myheadings{%
    \let\@oddfoot\@empty\let\@evenfoot\@empty
    \def\@evenhead{\slshape\leftmark\hfil}%       %EVEN PAGE
    \def\@oddhead{\hfil{\slshape\rightmark}}%     %ODD PAGE
    \let\@mkboth\@gobbletwo
    \let\sectionmark\@gobble
    \let\subsectionmark\@gobble
    }
\font\tenrm=cmr10
\font\tenit=cmti10 
\font\tenbf=cmbx10
\font\bfit=cmbxti10 at 10pt
\font\ninerm=cmr9
\font\nineit=cmti9
\font\ninebf=cmbx9
\font\eightrm=cmr8

\def\qed{\hbox{${\vcenter{\vbox{                    %HOLLOW SQUARE
   \hrule height 0.4pt\hbox{\vrule width 0.4pt height 6pt
   \kern5pt\vrule width 0.4pt}\hrule height 0.4pt}}}$}}

\renewcommand{\thefootnote}{\fnsymbol{footnote}}    %USE SYMBOLIC FOOTNOTE

\def\bsc{{\sc a\kern-6.4pt\sc a\kern-6.4pt\sc a}}       %LATEX LOGO
\def\bflatex{\bf L\kern-.30em\raise.3ex\hbox{\bsc}\kern-.14em 
T\kern-.1667em\lower.7ex\hbox{E}\kern-.125em X} 

\begin{document}
\setlength{\textheight}{7.7truein}  %for 2nd page onwards

\thispagestyle{empty}

%\markboth{\protect{\footnotesize\it Instructions for Typesetting
%Manuscripts}}{\protect{\footnotesize\it Instructions for
%Typesetting Manuscripts}}

\normalsize\textlineskip

\setcounter{page}{1}

\copyrightheading{}                     %{Vol. 0, No. 0 (1993) 000--000}

\vspace*{0.88truein}

%\fpage{1}
\centerline{\bf Extreme Long-time Dynamic Monte Carlo Simulations}
\vspace*{0.035truein}
\centerline{\bf for Metastable Decay in the $d$$=$$3$ Ising Ferromagnet}
\vspace*{0.37truein}
\centerline{\footnotesize Miroslav Kolesik
%\footnote{Typeset names in
%10 pt Times roman, uppercase. Use the footnote to indicate the
%present or permanent address of the author.}
}
\baselineskip=12pt
\centerline{\footnotesize\it Optical Sciences Center}
\baselineskip=10pt
\centerline{\footnotesize\it University of Arizona} 
\baselineskip=10pt
\centerline{\footnotesize\it Tucson, Arizona 85721, U.S.A.
%\footnote{State completely without abbreviations, the affiliation 
%and mailing address, including country. Typeset in 8 pt Times italic.}
}
\centerline{\footnotesize\it E-mail: kolesik@acms.arizona.edu}

%\vspace*{10pt}         %actual spacing
\vspace*{15pt}          %when needed
\centerline{\footnotesize M.A.\ Novotny}
\baselineskip=12pt
\centerline{\footnotesize\it Department of Physics and Astronomy}
\baselineskip=10pt
\centerline{\footnotesize\it Mississippi State University}
\baselineskip=10pt
\centerline{\footnotesize\it P.O.\ Box 5167}
\baselineskip=10pt
\centerline{\footnotesize\it Mississippi State, Mississippi, 
39762-5167, U.S.A. }
\centerline{\footnotesize\it E-mail: man40@ra.msstate.edu}

%\vspace*{10pt}         %actual spacing
\vspace*{15pt}          %when needed
\centerline{\footnotesize Per Arne Rikvold}
\baselineskip=12pt
\centerline{\footnotesize\it Department of Physics}
\baselineskip=10pt
\centerline{\footnotesize\it and Center for Materials Research and Technology}
\baselineskip=10pt
\centerline{\footnotesize\it and School of Computational Science 
and Information Technology}
\baselineskip=10pt
\centerline{\footnotesize\it Florida State University}
\baselineskip=10pt
\centerline{\footnotesize\it Tallahassee, Florida 32306-4350, U.S.A. }
\centerline{\footnotesize\it E-mail: rikvold@csit.fsu.edu}

\vspace*{0.225truein}
\publisher{(received date)}{(revised date)}

\vspace*{0.25truein}
\abstracts{
We study the extreme long-time behavior of the metastable 
phase of the three-dimensional Ising model with Glauber dynamics in an 
applied magnetic field and at a temperature below the critical 
temperature. For these simulations we use
the advanced simulation method of projective dynamics.  
The algorithm is described in detail, together with its application to the 
escape from the metastable state. 
Our results for the field dependence of the metastable lifetime are 
in good agreement with theoretical expectations and span 
more than {\it fifty decades\/} in time.
}{}{}

\vspace*{5pt}

%\textlineskip                  %) USE THIS MEASUREMENT WHEN THERE IS
%\vspace*{12pt}                 %) NO SECTION HEADING

\vspace*{1pt}\textlineskip      %) USE THIS MEASUREMENT WHEN THERE IS
\section{Introduction}         %) A SECTION HEADING
\vspace*{-0.5pt}
\noindent
One of the most difficult and most important problems in simulation in 
the sciences and engineering is that of disparate time and 
length scales.  The difficulty in the time-scale problem is the wide 
separation between the time scales of microscopic and macroscopic phenomena.  
Consider a simulation to study the time dependence of a nanoscale 
ferromagnetic particle or thin film using a dynamic Monte Carlo procedure.  
The microscopic time scale corresponds approximately to an inverse phonon 
frequency, or about $10^{-13}$~seconds.  The time over which simulations 
need to be performed corresponds to the time scale over which random 
thermal events change the direction of the magnetization.  For 
data recording applications this corresponds to the time over 
which data written on magnetic recording media should be stored: 
years to decades.  For studies in paleomagnetism, the time scales 
to be simulated are millions of years.  Bridging such disparate 
time scales obviously requires faster-than-real-time simulation algorithms.  

In this paper we present in detail a method to perform simulations 
that extend over extremely long 
times for the short-ranged ferromagnetic three-dimensional 
Ising model with Glauber dynamics at a temperature below the critical 
temperature. The method can be generalized to other dimensions, to other 
models with discrete state spaces, and to other dynamics.  The fact that 
many, many decades in time can be obtained in computer simulations of 
metastable decay of the $d=3$ Ising ferromagnet shows that the 
required faster-than-real-time simulations are indeed feasible.  

The remainder of this paper is organized as follows. 
The general features of metastable decay in kinetic Ising models are 
discussed in Sec.~2. 
The projective dynamics method of accelerated dynamic Monte Carlo simulation 
is briefly described in Sec.~3. 
Our numerical results are presented in Sec.~4, while Sec.~5 is devoted to 
a brief discussion and conclusions.

\setcounter{footnote}{0}
\renewcommand{\thefootnote}{\alph{footnote}}

\section{Metastability and the Ising Model}
\noindent

The Hamiltonian of the Ising model is given by 
\begin{equation}
{\cal H} = -J\sum_{\langle i,j\rangle} \sigma_i \sigma_j
- H \sum_i \sigma_i \;.
\label{Hamil}
\end{equation}
In Eq.~(\ref{Hamil}), $J$ is the nearest-neighbor exchange interaction 
which we take as ferromagnetic and for simplicity of notation set 
equal to unity, $H$ is the applied external magnetic field, 
the first sum is over all nearest-neighbor pairs of spins on a $d$-dimensional 
hypercubic lattice, and the second sum is over all 
$N=L^d$ spins.  Each spin can take two values, denoted by `up' and 
`down' or $\sigma=\pm1$. We impose periodic boundary conditions.  
(The projective dynamics method could also be used with other boundary 
conditions, but it is simplest for periodic boundary conditions.) 
The temperature is chosen below the critical temperature $T_c$ (in this 
work, 0.6$T_c$), so that in zero field the system has two degenerate, ordered 
phases with magnetizations near $\pm 1$. In an applied field, the ordered 
phase in which the spins are aligned opposite to the field becomes metastable. 
The value of the critical temperature used here corresponds to  
$J / k_{\rm B} T_c = 0.22165$, as obtained in high-precision Monte Carlo
simulations.\cite{BAIL92,TALA96}

The dynamic used corresponds to one derived from 
consideration of quantum spin~$1\over 2$ particles interacting with a 
fermionic heat bath.  In certain limits, it has been shown\cite{MART} that 
this corresponds to 
the dynamic: 1) a spin is chosen at random from among all $N$ spins; 
2) the spin is flipped with the Glauber\cite{GLAU} transition probability
\begin{equation}
p = {{\exp(-\beta E_{\rm new})}\over
{\exp(-\beta E_{\rm new})+\exp(-\beta E_{\rm old})}} 
= {1\over{1+\exp\left[\beta\left(E_{\rm new}-E_{\rm old}\right)\right]}} \;,
\label{eq:Glaub}
\end{equation}
where $\beta=1/(k_{\rm B} T)$ and $k_{\rm B}$ is Boltzmann's constant, 
the energy of the current configuration is $E_{\rm old}$, 
and the energy of the configuration obtained if the chosen spin is flipped 
is $E_{\rm new}$.  
One cycle through these two steps, whether or not a new configuration is 
obtained, is one Monte Carlo step.  Time is measured in units of $N$ such 
cycles, which is called one Monte Carlo step per spin (MCSS).  

Note that the dynamic described above corresponds to a 
particular physical dynamic, 
and that the time dependence of the system {\it with this dynamic\/} 
is the quantity 
of physical interest.  Consequently, the dynamic cannot be changed 
by the simulation algorithm.  
To implement a faster-than-real-time simulation the dynamic can only 
be implemented on a computer in a more intelligent fashion.  The 
restriction against changing the dynamic means that common 
advanced simulation techniques 
such as cluster algorithms, multicanonical methods, and simulated tempering, 
cannot be used since they all change the underlying dynamic of the system.  
Also note the restriction to randomly (rather than sequentially) 
picking the spin in the first step.  In some cases it has been shown that 
dynamic results differ between sequential and random updates.  
In particular, the prefactor in metastable decay has been shown in some cases 
to depend on whether the spin is chosen randomly or 
sequentially.\cite{RikGor,Prefactors}  
We use random updates for two reasons.  First and foremost 
this procedure corresponds to 
the dynamic obtained from coupling the quantum system to the heat bath.  
Second, advanced dynamic simulation algorithms, such as 
projective dynamics, are easier to implement 
in the random update scheme.  

We wish to study the lifetime of the metastable phase of the 
simple cubic Ising model with periodic boundary conditions.  
The system is initially prepared with all spins up (all $\sigma=1$), 
and the applied field $H$ is negative.  
Starting from this metastable initial state, we measure the metastable 
lifetime $\tau$ as the time 
when the magnetization first attains or crosses a pre-determined stopping 
value, $m_{\rm stop}$ ({\it i.e.\/}, the first-passage time to $m_{\rm stop}$).  
The particular value of $m_{\rm stop}$ is not very 
important, as long as it is chosen between the value corresponding to
the saddle point and the value corresponding to the equilibrium state, 
and $\tau$ does not have a substantial contribution from the time required to 
slide from the saddle point toward the equilibrium state. In this work,
we make the customary choice of $m_{\rm stop}=0$. 
The average lifetime, $\langle\tau\rangle$, is obtained by 
averaging $\tau$ over many realizations, each of which uses a different 
random number sequence.  The dependence of $\langle\tau\rangle$ on $T$ and $H$ 
is of central physical importance.  

\section{Projective Dynamics for the $d=3$ Ising Model}
\noindent

The slow-forcing method we apply to study the extremely long-lived 
metastable states was described in detail 
in Refs.~\cite{ProjBook,ProjLett,SlowForcing,MCReview}
For the sake of completeness, we describe briefly the main features of the 
approach here. 
It was designed for simulation of the escape from a 
metastable state with a very long lifetime. Its core is based
on an $n$-fold way simulation 
algorithm\cite{MCReview,nfoldBKL,nfoldMark,MCAMCMark} 
that preserves the
prescribed local Monte Carlo dynamic but eliminates all unsuccessful
attempted updates. This method is suitable for measuring lifetimes
many orders of magnitude longer than those accessible by conventional
algorithms. To extend the reach of simulations to even radically
longer lifetimes, the $n$-fold way method has to be augmented in two ways.
First, we have observed and proven that the mean lifetime of a metastable
state can be calculated from the so-called projected growth and shrinkage
rates of the stable phase within the metastable phase.\cite{SlowForcing}
To obtain an exact answer for the mean lifetime, one of course needs
to know the growth and shrinkage rates exactly as functions 
of the total volume fraction of the stable-phase regions. 
Since we do not have this exact information, we have to resort
to measuring the rates. It can be shown that the $n$-fold way simulation
itself samples these rates in an appropriate way and can thus be used
to obtain estimates of the growth and shrinkage rates as functions of
the total volume fraction of the stable phase.\cite{SlowForcing,MCReview}
However, although such an approach provides certain advantages, for example 
in studying the size dependence of the mean lifetime,\cite{ProjBook} 
it does not actually extend the regime of measurable
lifetimes. To achieve that, we make the second departure from the 
straightforward simulation approach. Namely, we modify the a-priori 
part of the Monte Carlo transition rates
in such a way that the volume of the stable phase is not allowed to
decrease below a certain value. This value is gradually increased 
during the simulation
to force the system from its metastable free-energy minimum over the
barrier into the stable phase. We call this minimal allowed volume
of the stable phase a forcing constraint in accordance with its effect on the
dynamics of the system. The simulation is initialized with the 
forcing constraint equal to zero and an $n$-fold way simulation 
is performed while the constraint is increased very slowly.
This way, the system samples 
configurations along the path of the escape from the metastable free-energy
minimum with the correct weights for measuring concentrations of spins 
in different classes. These spin classes and the corresponding energy 
changes are shown in Table~\ref{table1}. 
After the system reaches the states sufficiently
close to the top of the free-energy barrier and overcomes it,  it
converges quickly toward the true stable state. At this stage the
forcing constraint becomes irrelevant because of the relative
rapidity of the last escape stage.
Naturally, the duration of the forced escape has
nothing to do with the true metastable lifetime, because we have
modified the dynamic in a radical way. However, during the
multiple repetitions of the forced escapes, we gather sufficient statistics
to estimate the growth and shrinkage rates along the whole
escape path, which in turn are used to calculate an estimate of the 
{\it actual\/} mean lifetime. 
\begin{table}[htbp]
\tcaption{The 14 energy classes for the simple cubic Ising lattice.}
%\centerline{\footnotesize NP}
\centerline{\footnotesize\smalllineskip
\begin{tabular}{r r r l }\\
\hline
{} Class Number {} & Spin & Number of nn spins up  & 
$E_{\rm new}-E_{\rm old}$ \\
\hline
  1 &   $\uparrow$ & 6 & $-2H + 12J$ \\
  2 &   $\uparrow$ & 5 & $-2H + 8J$ \\
  3 &   $\uparrow$ & 4 & $-2H + 4J$ \\
  4 &   $\uparrow$ & 3 & $-2H$ \\
  5 &   $\uparrow$ & 2 & $-2H - 4J$ \\
  6 &   $\uparrow$ & 1 & $-2H - 8J$ \\
  7 &   $\uparrow$ & 0 & $-2H - 12J$ \\
  8 & $\downarrow$ & 6 & $+2H - 12J$ \\
  9 & $\downarrow$ & 5 & $+2H - 8J$ \\
 10 & $\downarrow$ & 4 & $+2H - 4J$ \\
 11 & $\downarrow$ & 3 & $+2H$ \\
 12 & $\downarrow$ & 2 & $+2H + 4J$ \\
 13 & $\downarrow$ & 1 & $+2H + 8J$ \\
 14 & $\downarrow$ & 0 & $+2H + 12J$ \\
\hline\\
\end{tabular}}
\label{table1}
\end{table}
Naturally, the question arises of whether the forced, modified dynamic
leaves the growth and shrinkage rates unaltered. As may be expected,
the rates and the corresponding mean lifetime actually do depend on the 
forcing speed. However, with decreasing forcing speed they converge
to their slow-forcing limit.\cite{SlowForcing,MCReview} This convergence
is sufficiently fast to obtain estimates of exceedingly long lifetimes.
While we have no formal proof for the convergence, it is possible to see
intuitively why it works.  
During the early stages of the forced escape, the sampled
configurations and values of the growth and shrinkage rates are close
to the states of the unperturbed escape provided the great majority of 
the Monte Carlo moves are not constrained. The sufficiently rare invocation
of the constrained dynamic rule can be achieved at
an appropriately slow forcing rate. 
Close to and behind the
free-energy barrier, the evolution is rapid, and the volume of the
stable phase grows faster than the progress of the forcing constraint.
Therefore, at this stage the system follows its natural non-equilibrium
path toward the stable phase. It is mainly this stage which requires 
one to repeat the forced escape many times in order to gather sufficient 
statistics for the non-equilibrium portion of the escape path.

Before discussing our numerical 
results in Sec.~4, we present a summary of the formulas needed
to interpret the simulation data. Again, the reader is referred to the
above references for details.
The growth and shrinkage rates in the slow-forcing limit are defined
in terms of the average class populations and Monte Carlo dynamic transition
rates. While the latter are given or specified by the simulated system
[Eq.~(\ref{eq:Glaub})],
the meaning of the former has to be made more accurate. These class
populations are normalized probabilities for the occurrence of
spins with the corresponding nearest-neighbor configurations. The average
is taken with respect to the ensemble generated by an unperturbed
escape from the metastable state.
The sampling during the repeated forced escapes provides us
with an estimate of the growth rates $g(n)$ and shrinkage rates $s(n)$
for a system with $n$ spins down, 
\begin{equation}
g(n)=\sum_{a=0}^6 c^{\uparrow  }_{a}(n)\  p^{\uparrow \downarrow}_{a}\ , \
s(n)=\sum_{a=0}^6 c^{\downarrow}_{a}(n)\  p^{\downarrow \uparrow}_{a} \;,
\label{eq:gsrates}
\end{equation}
where 
$p^{\sigma \sigma'}_{a}$
are the prescribed Monte Carlo spin-flip transition rates from $\sigma$ 
to $\sigma'$ for spins with $a$ nearest neighbors up, and $c^{\sigma}_{a}(n)$ 
stand for the spin-class populations which are sampled in the process.
The growth and shrinkage rates, $g(n)$ and $s(n)$, have to be obtained
for all $n$ between zero and 
$n_{\rm stop} \approx n_{\rm stable}$ that corresponds to
the stable-phase magnetization. As mentioned above, 
the precise setting of $n_{\rm stop}$ 
is irrelevant as long as it is well beyond the top of the free-energy barrier
separating the metastable phase from the stable phase.

The mean lifetime is then given by the formula
\begin{equation} 
\langle\tau\rangle = \sum_{n=0}^{n_{\rm stop}} h(n)\ \ , \ \
h(n-1) = { L^{-d} + s(n) h(n) \over g(n-1)} \ ,  
\label{eq:lifetime}
\end{equation} 
where $h(n)$ is the cumulative residence time in the state with $n$ spins down. 
As noted before, this expression is formally exact if the growth and shrinkage 
rates in the slow-forcing limit could be obtained. 
It represents a sum of the residence times $h(n)$ in the configurations with
$n$ spins down. The values of $h(n)$ make up a histogram
that is peaked sharply around the value of $n$ 
corresponding to the magnetization 
of the metastable phase. Our approximation consists in 
replacing $g(n)$ and $s(n)$ by their corresponding estimates
obtained at a finite forcing rate.

\section{Results}
\noindent

We have performed a series of simulations using both the direct
$n$-fold method and the forced-escape method to obtain the mean lifetime
of the metastable state of the three-dimensional ferromagnetic Ising 
model with Glauber dynamic at a temperature of 0.6$T_c$ (this is about
11\% above the roughening temperature\cite{HASE94}). 
Below we present results 
for the system size of $L^d = 16^3$ and a range of external fields.

The data acquired by the forced-escape method for a set of
chosen strengths of the external field were utilized to
obtain estimates of the spin-class populations and
of the growth and shrinkage rates at other values of the external field
by straightforward interpolation and extrapolation.

\begin{figure}[htbp] %ORIGINAL SIZE: width=1.4TRUEIN; height=1.5TRUEIN
\vspace*{13pt}
%\centerline{\epsfig{scale=1.0,file=obrtau1.eps} } 
\centerline{\epsfig{scale=1.0,file=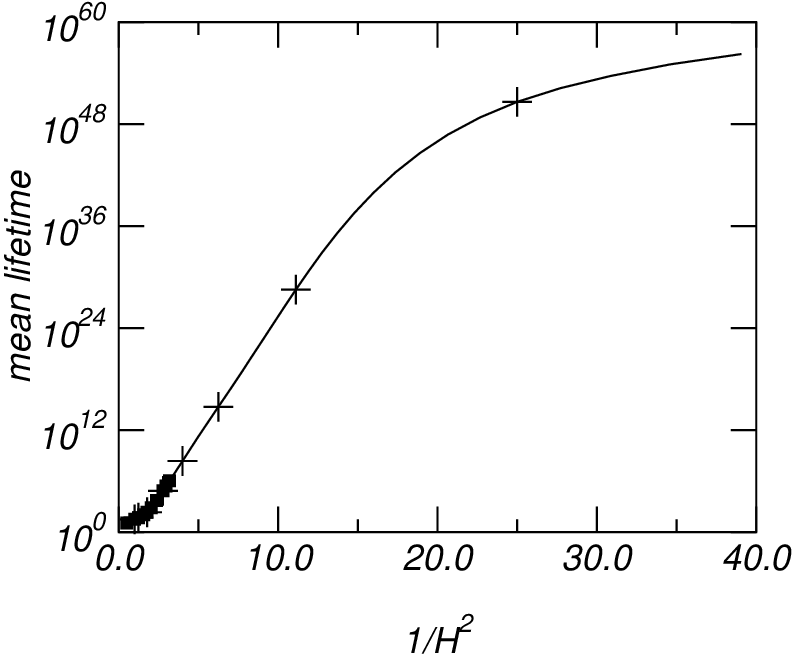} } 
\vspace*{13pt}
\fcaption{
\label{fig:whole}
The average lifetime $\langle\tau\rangle$ 
in units of Monte Carlo steps per spin (MCSS) for the 
$d=3$ ferromagnetic Ising model with Glauber dynamics at a temperature 
of $0.6T_c$. The lifetime is shown on a logarithmic scale as a function of 
$1/H^2$.  Note the 
{\it extremely } long lifetimes obtained in this computer simulation.  
The cross symbols indicate external field values for which
the growth and shrinkage rates were measured, while the full curve was
calculated from interpolated and extrapolated rates.
The filled squares in the lower left-hand (strong field, short time) 
corner of the figure represent standard $n$-fold way simulations. 
A magnified view of this region is shown in Fig.~\protect\ref{fig:part}. 
}
\end{figure}

Figure~\ref{fig:whole} shows the global picture of the mean lifetime
as a function of the external field that drives
the system toward its true equilibrium state. 
The lifetime data are plotted on a logarithmic scale versus 
$1/|H|^{d-1} = 1/H^2$, so that data in both the single-droplet and
multidroplet decay regimes should show up as approximately straight lines. 
(Different regimes of the magnetization reversal and the rationale for
this plotting method are described 
in Refs.~\cite{RikGor,Prefactors,MMM,HowJAP,HowPRB,HowBC})
The straight portion of the graph in Fig.~\ref{fig:whole}
corresponds to the single-droplet
regime, in which the escape from metastability is triggered by
a single critical droplet of the stable phase.
The bend in the strong-field portion of the curve
is the cross-over to the multidroplet regime, in which many
supercritical droplets exist by the time the system reaches the 
cut-off magnetization
(the corresponding field is 
known as "the dynamic spinodal"\,\cite{RikGor,Prefactors}). 
The gradual bend in the
weak-field region is due to the cross-over into the coexistence regime,
which is characterized by a critical fluctuation comparable
in size with the system itself. 

\begin{figure}[htbp] %ORIGINAL SIZE: width=1.4TRUEIN; height=1.5TRUEIN
\vspace*{13pt}
%\centerline{\epsfig{scale=1.0,file=obrtau2.eps} }  
\centerline{\epsfig{scale=1.0,file=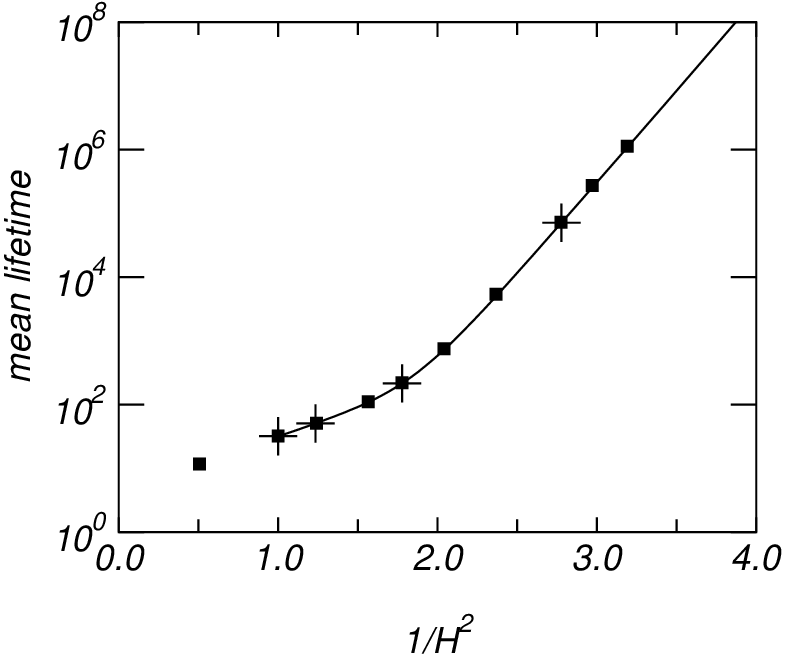} }  
\vspace*{13pt}
\fcaption{
\label{fig:part}
The magnified region of the relatively short lifetimes from 
Fig.~\protect\ref{fig:whole}. 
In the lower part of this region, direct $n$-fold way 
simulations are feasible. The corresponding direct simulation estimates
are indicated by filled squares. 
As in Fig.~\protect\ref{fig:whole}, 
the crosses represent slow-forcing simulations, and the solid curve was 
calculated from interpolated rates. 
}
\end{figure}

Note the exceedingly long lifetimes our approach enables us to estimate. 
(To put the range of time scales into perspective, we note that the age of 
the universe, measured in femtoseconds, is about $10^{33}$.) 
By itself, the fact that we can reach almost to the coexistence regime
demonstrates the power of the method. Standard $n$-fold way
simulation results are also included in the figure to contrast the 
corresponding orders of magnitude. To make the details at stronger
fields discernible, we present in Fig.~\ref{fig:part} 
a blow-up of the lower-left portion of Fig.~\ref{fig:whole}. 
There one can see that direct lifetime
measurements are perfectly reproduced by the slow-forcing 
escape simulations, even though the latter were performed only for a limited
set of external field values.

Naturally, with estimated lifetimes this long, and in the absence of
direct comparison with other computational methods, the question arises
of how accurate our estimates actually are. This is, of course, a 
difficult question to answer. One way to approach this problem
is to look at the theoretically predicted behavior of the mean lifetime
as a function of the external field strength. Namely,
the dominant field dependence is expected to be universal,
having the form of an exponential modified by a field-dependent
prefactor. The droplet theory of metastable decay 
predicts a characteristic behavior for
the effective slopes of the logarithm of the mean 
lifetime as plotted in Figs.~\ref{fig:whole} 
and~\ref{fig:part},\cite{RikGor,Prefactors} 
\begin{equation}
\Lambda_{\rm eff} = \frac{{\rm d} \ln{\langle\tau\rangle}}{{\rm d} (1/|H|^{d-1})}
                  = \lambda |H|^{d-1} + \Lambda
\;,
\end{equation}
where the values of $\lambda$ and $\Lambda$ depend on
the magnetization reversal regime [single-droplet (SD) or multidroplet (MD)]
and on the microscopic dynamic. 
In three dimensions and with a dynamic with updates at randomly chosen
sites as used here, field-theoretical calculations\cite{GNW} predict for 
the single-droplet 
regime $\lambda_{\rm SD} = -1/6$ and $\Lambda_{\rm SD} = \beta \Xi(T)$, 
where $\Xi(T)$ is the field-independent part of the free energy of the 
critical droplet.\cite{RikGor,Prefactors} 
In the multidroplet regime, the field-theoretical results, combined with the 
Kolmogorov-Johnson-Mehl-Avrami theory of metastable decay in large 
systems,\cite{RikGor,KJMA,Ramos} yield $\lambda_{\rm MD} = +1/3$ and 
$\Lambda_{\rm MD} = \beta \Xi(T)/(d+1) = \beta \Xi(T)/4$.  
From the droplet theory one obtains $\Xi(T)$ (here given
specifically for three dimensions) as  
$\Xi(T) \approx \Omega_3 \sigma^3 / m^2$.\cite{RikGor,Prefactors} 
Here $\sigma(T)$ is the equilibrium surface tension of an interface 
parallel to one of the lattice symmetry directions, 
separating the positively and negatively magnetized phases, 
and $m(T)$ is the spontaneous equilibrium
magnetization in zero field. The quantity $\Omega_3(T)$ is a shape
factor which interpolates smoothly between 8 for cubic droplets in the
low-temperature limit and 4$\pi$/3 for spherical droplets at higher
temperatures, and it could in principle be obtained by a Wulff
construction with the full, anisotropic surface tension.\cite{RikGor,CCAG94} 
Since the latter is not available, we simply use the extreme
values of $\Omega_3$ to obtain lower and upper bounds on $\beta \Xi(0.6T_c)$. 
For this purpose we use estimates for the surface tension and
magnetization obtained from high-precision equilibrium Monte Carlo
simulations, $\sigma(0.6T_c)/J \approx 1.536$,\cite{HASE94} and 
$m(0.6T_c) \approx 0.971$.\cite{TALATOO} 
The resulting approximate lower and upper bounds on 
$\beta \Xi(0.6T_c)$ are 5.948 and 11.359, respectively. 

\begin{figure}[htbp] %ORIGINAL SIZE: width=1.4TRUEIN; height=1.5TRUEIN
\vspace*{13pt}
%\centerline{\epsfig{scale=0.8,file=NEWdlntaudHm2.eps} }  
\centerline{\epsfig{scale=0.8,file=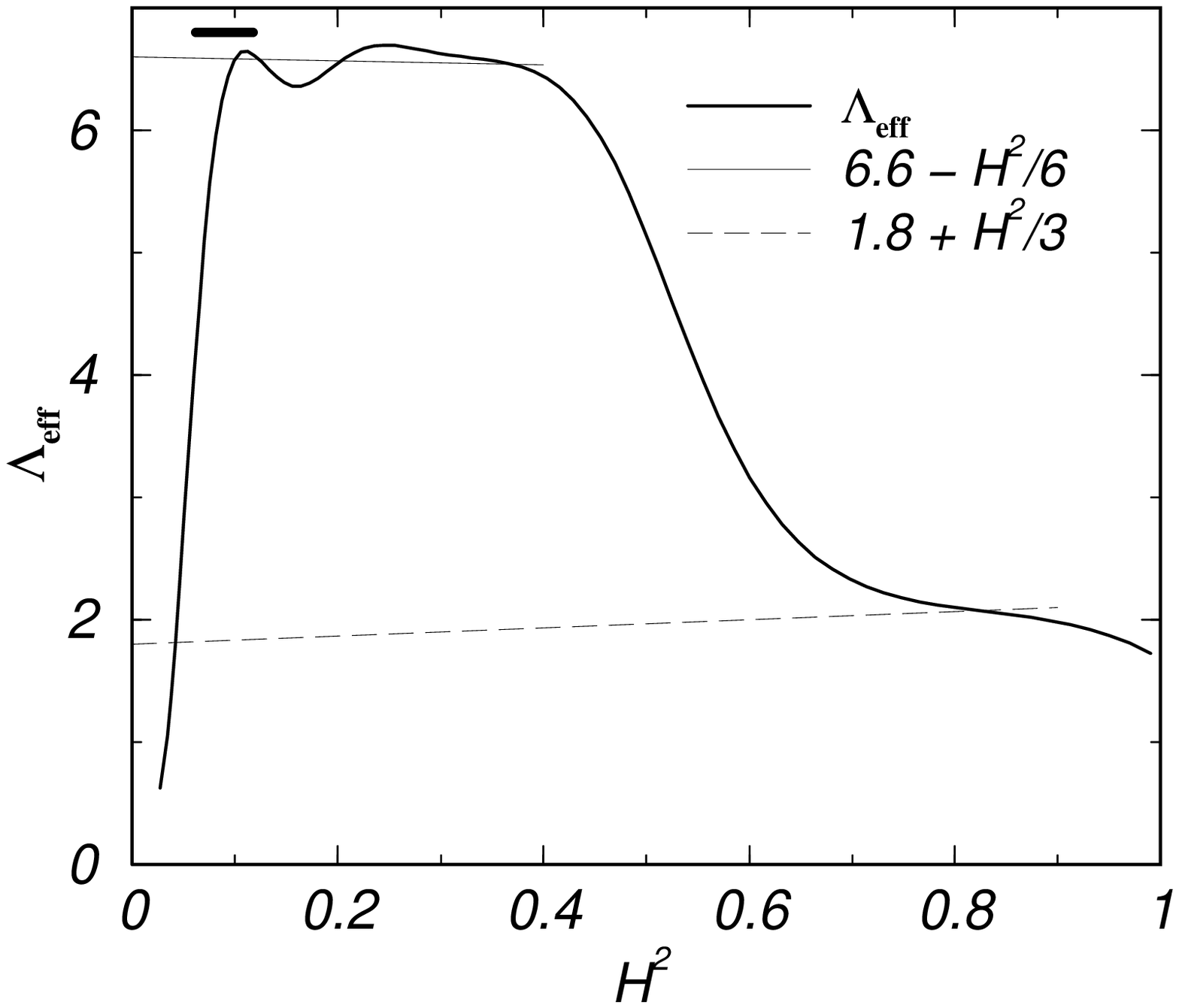} }  
\vspace*{13pt}
\fcaption{
\label{fig:slopes}
Numerical estimate of the effective slope $\Lambda_{\rm eff}$, shown vs $H^2$.
The thick solid curve is obtained by numerical differentiation of 
the curve shown in Fig.~\protect\ref{fig:whole}. 
The thin solid and dashed straight lines are 
theoretical predictions for the single-droplet and multidroplet regimes, 
respectively, with $y$-intercepts fitted as described in the text. 
The horizontal bar in the upper left corner represents the theoretical
range of $H^2$ for the cross-over to the coexistence regime. 
}
\end{figure}

The quantities that enter into $\Lambda_{\rm eff}$ are very sensitive
to sampling errors, and they are not easy to obtain, even from direct simulations
when such are feasible. It is therefore a good test of our results
to evaluate the effective slopes to see whether the expected characteristic
behaviors are recovered. In Fig.~\ref{fig:slopes}  
we have plotted $\Lambda_{\rm eff}$ vs $|H|^{d-1} = H^2$, where
the continuous curve is obtained by numerical differentiation of the  
mean lifetimes, which in turn were calculated from  the
interpolated and extrapolated class-population data.
We see that the main features of the expected cross-over between
the multidroplet and single-droplet regimes is well pronounced,  
as is the rapid decrease of $\Lambda_{\rm eff}$ 
for very weak fields that corresponds 
to the cross-over to the coexistence regime (see below).
However, our effective slope curve exhibits modulations which are
clearly artifacts of the interpolation procedure. These modulations prevent 
us from determining the simulated values of $\lambda_{\rm SD}$ and 
$\lambda_{\rm MD}$. Instead, we have placed straight lines with the 
theoretically predicted values of $\lambda_{\rm SD}$ and $\lambda_{\rm MD}$, 
so as to independently 
match $\Lambda_{\rm eff}$ in the single-droplet and 
multidroplet regimes reasonably by eye. The resulting
intercepts with the vertical axis give estimates 
$\Lambda_{\rm SD} = \beta \Xi(0.6T_c) \approx 6.6$ 
and $\Lambda_{\rm MD} \approx 1.8$. Thus, our estimate for $\beta \Xi(0.6T_c)$
lies between the lower and upper bounds obtained from droplet theory. As
it lies closer to the lower bound, our simulations may indicate that the
average critical droplet is closer to spherical than to cubic.  
Furthermore, the ratio $\Lambda_{\rm SD}/\Lambda_{\rm MD} \approx 3.7$,
in reasonable agreement with its expected value of 4. 

Another characteristic feature of Fig.~\ref{fig:slopes} is the steep drop
in $\Lambda_{\rm eff}$ for weak fields. At the field where this drop sets in
(known as "the thermodynamic spinodal field",\cite{RikGor,Prefactors}
$H_{\rm THSP}$), the free energy of the critical droplet 
equals that of a slab of the same volume, which spans 
the system in two dimensions. 
This field can also be obtained analytically by droplet-theoretical
arguments \cite{LEUN90,JLEE95} as (given specifically for three dimensions)
$H_{\rm THSP} \approx \sigma \sqrt{3 \Omega_3 / 2} / (mL)$, yielding 
lower and upper bounds at $T=0.6T_c$ of 0.2478 and 0.3425, respectively. 
The corresponding range for $H^2$ is shown in Fig.~\ref{fig:slopes} as a
thick, horizontal bar above the $\Lambda_{\rm eff}$ curve. The agreement
between our simulation results and this theoretical prediction is also very good. 

The above considerations give us
a feeling for how accurate the forced-escape lifetime estimates are:
the method provides sufficiently reliable results for the lifetimes
themselves, but the coarse sampling of the external field 
used in the present work 
prevents us from taking numerical derivatives of the lifetime
with sufficient accuracy to reliably 
estimate their dependence on the applied field. 
To improve the effective-slope estimates, it would
not only be necessary to perform the forced-escape simulations over
a denser set of field strengths, but also to increase the number of
simulated escapes at each field strength. To increase the accuracy of the 
slope determination in the multidroplet regime, it would also be necessary 
to use larger systems to increase the range of fields corresponding to 
that regime. Nevertheless, we find the agreement with the theoretical predictions 
of our estimates for $\Lambda_{\rm SD}$, $\Lambda_{\rm MD}$, and the
cross-over field to the weak-field coexistence regime quite encouraging. 

%\clearpage
\section{Discussion and Conclusions}
\noindent

We have described the method of projective dynamics with 
slow forcing and applied it to the long-time simulation 
of the metastable lifetime of 
the simple-cubic Ising model with the Glauber dynamic.  
We have obtained the lifetime corresponding to the dynamic 
Monte Carlo simulation over 
more than {\it fifty orders of magnitude\/} without changing the dynamics 
of the model!  From the dependence of the lifetime on the 
applied field we are able to identify different regimes of decay for 
the model.\cite{RikGor,Prefactors}  The results thus far are good enough to 
see the expected dependence of the lifetime on the applied field.  
This slow-forcing projective dynamics method can also be used in 
other simulations of discrete-state 
models to bridge widely disparate time scales.  

\nonumsection{Acknowledgments}
\noindent
Useful discussions with S.J.\ Mitchell are acknowledged.  
This research was funded partly the the U.S.\ National Science 
Foundation through grants No.\ DMR-9871455 and DMR-0120310.  
M.K.\ was partly supported by the Slovak Grant Agency through
Grant No.\ 2/7174/20.

\newpage

\begin{thebibliography}{000}

\bibitem{BAIL92}
C.F.\ Baillie, R.~Gupta, K.A.\ Hawick, and G.S.\ Pawley, 
{\bibit Phys.\ Rev. B\/} {\bibbf 45}, 10438 (1992). 

\bibitem{TALA96}
A.L.\ Talapov and H.W.J.\ Bl{\"o}te, 
{\bibit J.\ Phys.\ A: Math.\ Gen.\/} {\bibbf 29}, 5727 (1996). 

\bibitem{MART}
Ph.A.\ Martin, {\bibit J.\ Stat.\ Phys.\/} {\bibbf 16}, 149 (1977). 

\bibitem{GLAU}
R.J.\ Glauber, {\bibit J.\ Math.\ Phys.\/} {\bibbf 4}, 294 (1963). 

\bibitem{RikGor} 
P.A.\ Rikvold and B.M.\ Gorman, in 
{\bibit Annual Reviews of Computational Physics I}, 
edited by D.~Stauffer (World Scientific, Singapore, 1994), 
p.~149, and references therein.  

\bibitem{Prefactors}
P.A.\ Rikvold, H.~Tomita, S.~Miyashita, and S.W.\ Sides,
{\bibit Phys.\ Rev.\ E} {\bibbf 49}, 5080 (1994).

\bibitem{ProjBook}
M.~Kolesik, M.A.\ Novotny, P.A.\ Rikvold, and D.M.\ Townsley, in:
 {\bibit Computer Simulation Studies in Condensed Matter Physics X},
 edited by D.P.\ Landau, K.K.\ Mon, and H.-B.\ Sch{\"u}ttler 
 (Springer, Berlin, 1998) p.~246.

\bibitem{ProjLett}
M.~Kolesik, M.A.\ Novotny, and P.A.\ Rikvold,
 {\bibit Phys.\ Rev.\ Lett.\/} {\bibbf 80}, 3384 (1998).

\bibitem{SlowForcing}
M.A.\ Novotny, M.~Kolesik,  and P.A.\ Rikvold,
 {\bibit Comput.\ Phys.\ Commun.\/} {\bibbf 121-122}, 330 (1999).

\bibitem{MCReview}
M.A.\ Novotny, in 
{\bibit Annual Reviews of Computational Physics IX}, 
edited by D.~Stauffer (World Scientific, Singapore, 2001), 
p.~153.  

\bibitem{nfoldBKL}
A.B.\ Bortz, M.H.\ Kalos, and J.L.\ Lebowitz,
 {\bibit J.\ Comput.\ Phys.\/} {\bibbf 17}, 10 (1975).

\bibitem{nfoldMark}
M.A.\ Novotny, Computers in Physics {\bibbf 9}, 26 (1995).  

\bibitem{MCAMCMark} 
M.A.\ Novotny,
{\bibit Phys.\ Rev.\ Lett.\/} {\bibbf 74}, 1    (1995); 
                  Erratum {\bibbf 75}, 1424 (1995).

\bibitem{MMM}
H.L.\ Richards, S.W.\ Sides, M.A.\ Novotny, and P.A.\ Rikvold,
{\bibit J.\ Magn.\ Magn.\ Mater.\/} {\bibbf 150}, 37 (1995).

\bibitem{HowJAP}
H.L.\ Richards, S.W.\ Sides, M.A.\ Novotny, and P.A.\ Rikvold,
{\bibit J.\ Appl.\ Phys.\/} {\bibbf 79}, 5479 (1996).

\bibitem{HowPRB}
H.L.\ Richards, M.A.\ Novotny, and P.A.\ Rikvold,
{\bibit Phys.\ Rev.\ B} {\bibbf 54}, 4113 (1996).

\bibitem{HowBC}
H.L.\ Richards, M.\ Kolesik, P.-A. Lindg{\aa}rd, 
P.A.\ Rikvold, and M.A.\ Novotny, 
{\bibit Phys.\ Rev.\ B} {\bibbf 55}, 11521 (1997).

\bibitem{GNW}
N.J.\ G{\"u}nther, D.A.\ Nicole, and D.J.\ Wallace, 
{\bibit J.\ Phys.\ A} {\bibbf 13}, 1755 (1980). 

\bibitem{KJMA}
A.N.\ Kolmogorov, {\bibit Bull.\ Acad.\ Sci.\ USSR, Phys.\ Ser.\/} 
{\bibbf 1}, 355 (1937); 
W.A.\ Johnson and R.F.\ Mehl, 
{\bibit Trans.\ Am.\ Inst.\ Mining and Metallurgical Engineers\/} 
{\bibbf 135}, 416 (1939); 
M.~Avrami, {\bibit J.\ Chem.\ Phys.\/} {\bibbf 7}, 1103 (1939);
M.~Avrami, {\bibit J.\ Chem.\ Phys.\/} {\bibbf 8}, 212 (1940);
M.~Avrami, {\bibit J.\ Chem.\ Phys.\/} {\bibbf 9}, 177 (1941).

\bibitem{Ramos}
R.A.\ Ramos, P.A.\ Rikvold, and M.A.\ Novotny, 
{\bibit Phys.\ Rev.\ B\/} {\bibbf 59}, 9053 (1999).

\bibitem{CCAG94}
C.C.A.\ G{\"u}nther, P.A.\ Rikvold, and M.A.\ Novotny, 
{\bibit Physica A\/} {\bibbf 212}, 194 (1994). 

\bibitem{HASE94}
M.~Hasenbusch and K.~Pinn, {\bibit Physica A\/} {\bibbf 203}, 189 (1994). 
Our value for $\sigma(0.6Tc)$ was obtained by linear interpolation from
Table~VII of this reference. (Note that our $\sigma$ corresponds to
these authors' $\sigma/\beta$.) 

\bibitem{TALATOO}
Our value for $m(0.6T_c)$ was obtained from Fig.~1 of 
Ref.~\protect\cite{TALA96}, together with Eq.~(10) from the same reference. 

\bibitem{LEUN90}
K.~Leung and R.K.P.\ Zia, {\bibit J.\ Phys.\ A\/} {\bibbf 23}, 4593 (1990). 

\bibitem{JLEE95}
J.~Lee, M.A.\ Novotny, and P.A.\ Rikvold,
{\bibit Phys.\ Rev.\ E\/} {\bibbf 52}, 356 (1995). 

\end{thebibliography}
\end{document}